\documentclass[aps,pre,twocolumn]{revtex4}%
\usepackage{amsfonts}
\usepackage{amsmath}
\usepackage{amssymb}
\usepackage{graphicx}

\begin{document}

\title{Scaling theory of DNA confined in nanochannels and nanoslits}

\author{Theo Odijk*}

\affiliation{Complex Fluids Theory, Faculty of Applied Sciences, Delft
University of Technology, 2628 BC Delft, the Netherlands}

\begin{abstract}

A scaling analysis is presented of the statistics of long DNA confined in
nanochannels and nanoslits. It is argued that there are several regimes in
between the de Gennes and Odijk limits introduced long ago. The DNA chain
folds back on itself giving rise to a global persistence length which may be
very large owing to entropic deflection. Moreover, there is an orientational
excluded-volume effect between the DNA segments imposed solely by the
nanoconfinement. These two effects cause the chain statistics to be
intricate leading to nontrivial power laws for the chain extension in the
intermediate regimes. It is stressed that DNA confinement within nanochannels differs from that in nanoslits because the respective orientational excluded-volume
effects are not the same.

\vspace{20 pt}

*Electronic address: odijktcf@wanadoo.nl

\end{abstract}

\maketitle

A perusal of the rapidly developing literature on nanoconfined DNA shows
that its behavior is more complex than anticipated (see e.g.
\cite%
{TEG,REC,REI1,FAN,MAN,BAL1,STE,KIM,HSI,TAN,BAL2,JO,REI2,KRI,CRO,BON,KRI2}).
It appears that more regimes are needed besides those
originally described by Daoud and de Gennes \cite{DAO} and Odijk \cite{ODI1}%
. The nanoconfinement of a semiflexible chain specifically introduces
subtleties in the chain statistics which I address here within a scaling
analysis. A complete theory would involve solving a Fokker-Planck equation
subject to the boundary conditions arising from nanoconfinement \cite{SEM}%
. Nevertheless, backfolding or hairpin formation may
be addressed in a mechanical approximation \cite{ODI2} though entropic
depletion of the chain near a wall still needs to be
resolved quantitatively \cite{ODI3}. Numerical investigations of
nanoconfined stiff chains interacting via excluded-volume interactions have
appeared recently \cite{CHE1,CHE2,CHE3} but in the limit of ground-state
dominance without accounting for hairpin formation.

Let us first consider a very long double-stranded DNA chain confined in a
nanochannel of square cross section whose side $D$ is smaller than the
persistence length $P$ so that we are in the Odijk regime $(D<P)$. Thus the
chain may be conveniently viewed as a sequence of deflection segments of
typical length \cite{ODI1}

\begin{equation}
\lambda \simeq D^{\frac{2}{3}}P^{\frac{1}{3}}  \label{VGL1}
\end{equation}%
The orientational fluctuations with respect to the channel center axis are
given by the mean-square average

\begin{equation}
\left\langle \theta ^{2}\right\rangle \simeq c_{1}\left( \frac{D}{P}\right)
^{\frac{2}{3}}  \label{VGL2}
\end{equation}%
It is important to note that the coefficient $c_{1}$ here is quite small as
has been determined numerically \cite{WAN,YAN,WAG} and estimated
analytically \cite{JO}. Inevitably, a long chain must bear thermally
activated hairpins leading to a global persistence length $g$ as shown in
fig 1. The entropic depletion caused by the nanowalls forces the hairpin
bends to be tightened up so that $g$ is often considerably larger than the
persistence length $P$ \cite{ODI2}. Expressions for $g$ are presented in
appendix I.

\begin{figure}[htb!]
 \begin{center}
   \includegraphics[angle=270,scale=0.4]{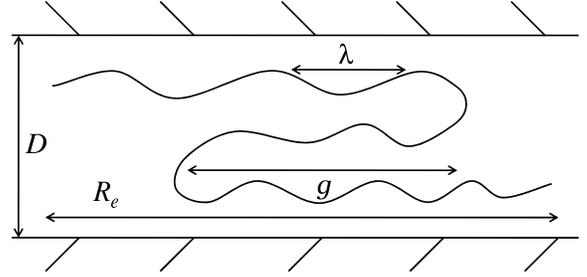}
   \caption{A DNA chain enclosed in a nanochannel of width $D$. The distance
between hairpins is typically $g$.\label{fig1}}
 \end{center}
\end{figure}

In view of the backfolding, segments of the DNA interact with each other via
the excluded-volume effect. Owing to the charges borne by the DNA backbone,
one introduces an effective diameter $d_{eff}$ rather than a bare diameter
\cite{FIX} ($P\gg d_{eff}$). If the interaction were purely isotropic, the excluded volume
between a pair of deflection segments would scale as \cite{ODI4}

\begin{equation}
\beta _{\lambda }\simeq \lambda ^{2}d_{eff}  \label{VGL3}
\end{equation}%
But the segments are aligned (see eq (\ref{VGL2})) so that the effective excluded
volume becomes \cite{ONS,KHO,ODI5}

\begin{equation}
E=\beta _{\lambda }\left\langle \left\vert \sin \delta \right\vert
\right\rangle  \label{VGL4}
\end{equation}

\begin{equation}
\left\langle \left\vert \sin \delta \right\vert \right\rangle \simeq \left(
\frac{D}{P}\right) ^{\frac{1}{3}}  \label{VGL5}
\end{equation}%
where $\delta $ is the angle between two deflection rodlets (for a
computation of the orientational factor, see appendix II). It is stressed
that we are in the sparse limit $d_{eff}\ll D$: the orientational order is
imposed solely by the walls of the nanochannel and independent of the
density of DNA segments.

The DNA chain of length $L$ may now be viewed as a one dimensional walk
consisting of $L/g$ statistical segments. It is partly self avoiding in the sense that the volume exclusion between the deflection rodlets is three-dimensional. Employing a mean-field argument of the Flory type which is
excellent in one dimension \cite{GEN1}, I write the free energy of the
confined chain as

\begin{equation}
\frac{F}{k_{B}T}\simeq \frac{R_{e}^{2}}{Lg}+\frac{N_{\lambda }^{2}E}{%
R_{e}D^{2}}  \label{VGL6}
\end{equation}%
Here, T is the temperature and $k_{B}$ is Boltzmann's constant. The first
term in eq (\ref{VGL6}) is the ideal free energy needed to extend the DNA chain to
root-mean-square extension $R_{e}$ and the second term arises from the
interaction of $N_{\lambda }=L/\lambda $ deflection segments in a
volume $R_{e}D^{2}$. Minimization of $F$ with respect to $R_{e}$ yields

\begin{equation}
R_{e}\simeq L\xi _{1}^{\frac{1}{3}}  \label{VGL7}
\end{equation}

\begin{equation}
\xi _{1}\equiv \frac{gE}{\lambda ^{2}D^{2}}\simeq \frac{gd_{eff}}{D^{\frac{5%
}{3}}P^{\frac{1}{3}}}  \label{VGL8}
\end{equation}%
Eqs (\ref{VGL6}) and (\ref{VGL7}) are reminiscent of those occurring in the
theory of grafted polymers \cite{ALE,GEN2}. One is now naturally led to introduce
the following regimes.

\subsection*{Regime 1: $\xi _{1}> 1$}

From eq (\ref{VGL7}) we discern that the chain must be almost fully aligned $\left( R_{e}\simeq L\right) $ apart from minor fluctuations given by eq (\ref{VGL2}).

\subsection*{Regime 2A: $\xi _{1}< 1$ provided $L> L_{\ast}$}

The excluded-volume term in eq (\ref{VGL6}) is analogous to the
excluded-volume parameter $Z$ introduced in the two-parameter theory of the
expansion of flexible polymer chains \cite{YAM1}

\begin{equation}
Z\simeq \frac{N_{\lambda }^{2}E}{D^{2}R_{e}}  \label{VGL9}
\end{equation}%
The excluded-volume effect is fully exerted in the limit $Z\gg 1$. There is
a crossover to the case of weakly interacting segments at $Z=O\left(
1\right) $. Hence, the contour length must be larger than $L_{\ast }$ if eq %
(\ref{VGL7}) is to remain valid.

\begin{equation}
L_{\ast }=g\xi _{1}^{-\frac{2}{3}}\simeq g^{\frac{1}{3}}D^{\frac{10}{9}}P^{%
\frac{2}{9}}d_{eff}^{-\frac{2}{3}}  \label{VGL10}
\end{equation}

\subsection*{Regime 2B: $L< L_{\ast}$}

In this case the excluded-volume effect is weak $\left( Z\ll 1\right) $ so
the behavior of the DNA is effectively that of an ideal chain as long as $%
L\gg g$

\begin{equation}
R_{e}^{2}\simeq Lg  \label{VGL11}
\end{equation}%
I now investigate what happens as the nanochannel is widened. The global
persistence length rapidly approaches the usual persistence length (see eq (%
\ref{VGL25}) in appendix I). Accordingly, one may introduce a crossover

\begin{equation}
D_{\ast }=c_{2}P  \label{VGL12}
\end{equation}%
at $g\simeq P$ which signals the breakdown of the Odijk regime (the
numerical coefficient $c_{2}$ is larger than unity). Concomitantly, the
channel is no longer narrow enough to impose orientational order on the DNA:
$\left\langle \left\vert \sin \delta \right\vert \right\rangle =O\left(
1\right) $ and $E\simeq P^{2}d_{eff}$. Thus, we now enter the next regime
upon increasing $D$.

\subsection*{Regime 3: $D_{\ast \ast }> D> D_{\ast }$}

I now express the total free energy of the chain as in eq (\ref{VGL6}) but
with $g=P$. This leads to

\begin{equation}
R_{e}\simeq L\xi _{2}^{\frac{1}{3}}  \label{VGL13}
\end{equation}

\begin{equation}
\xi _{2}\equiv \frac{Pd_{eff}}{D^{2}}  \label{VGL14}
\end{equation}%
It is again possible to demarcate one subregime in which the chain expansion
is dominated by the excluded-volume effect from another subregime where the
chain is ideal more or less. The crossover in the contour length is given by

\begin{equation}
L_{\ast }\simeq \frac{P^{\frac{1}{3}}D^{\frac{4}{3}}}{d_{eff}^{\frac{2}{3}}}
\label{VGL15}
\end{equation}%
Superficially, it may appear as if eq (\ref{VGL13}) conforms to a Daoud-de
Gennes type of theory \cite{DAO} but this is not the case for $D< D_{\ast \ast }$ (see eq (\ref{VGL16})). The intermediate regime one has to introduce here is
caused by the fact that the DNA segments are slender $\left( d_{eff}\ll
P\right) $. The chain may be viewed as a sequence of anisometric blobs, each of length $(L_{\ast}P)^\frac{1}{2}$ and diameter $D$.

\subsection*{Regime 4: $D> D_{\ast \ast }$}

Daoud and de Gennes argued that a flexible polymer confined in a capillary
piles up as a sequence of blobs, each blob being viewed as a Flory chain of $%
m$ segments \cite{DAO}. The blobs do not interpenetrate owing to the excluded-volume
repulsion. The supposition is that $Z\gg 1$ within a blob. In the problem at
hand, we have $m$ segments of length $P$ interacting by an excluded volume $%
P^{2}d_{eff}$ yielding an excluded-volume parameter $Z=m^{2}\beta /m^{\frac{3%
}{2}}P^{3}=m^{\frac{1}{2}}\left( d_{eff}/P\right) $. If the Flory expansion
is to be valid within a blob of radius $D$, we require $D\simeq \left( m^{%
\frac{1}{2}}P\right) Z^{\frac{1}{5}}$. In other words, upon eliminating $m$
we must have $Z=\left( Dd_{eff}/P^{2}\right) ^{\frac{5}{6}}$. Therefore, if
the blob picture is to be valid, one has to impose $Z> 1$ implying that $D_{\ast \ast }$ is expressed by

\begin{equation}
D_{\ast \ast }\equiv P^{2}/d_{eff}  \label{VGL16}
\end{equation}%
Note that eq (\ref{VGL13}) remains valid as can be verified in a blob
analysis. However, the difference between regimes 3 and 4 may show up in
subtle measurements.

Let us next turn to nanoslits of rectangular cross section $A\times D$ ($A>
D$). In many respects, the reasoning is now the same as that
presented above so the analysis will be brief. I first focus on thin slits $%
\left( D\leq \pi P\right) $ in which the DNA chain is effectively one
dimensional.

The analogue of eq (\ref{VGL6}) is now

\begin{equation}
\frac{F}{k_{B}T}\simeq \frac{R_{e}^{2}}{gL}+\frac{N_{\lambda }^{2}E}{R_{e}AD}
\label{VGL17}
\end{equation}%
The orientational factor $\left\langle \left\vert \sin \delta \right\vert
\right\rangle $ within E then has a rather subtle dependence on A and D (see
appendix II). The deflection length is still given by eq (\ref{VGL1}). Upon
minimizing eq (\ref{VGL17}), we get

\begin{equation}
R_{e}\simeq L\xi _{3}^{\frac{1}{3}}  \label{VGL18}
\end{equation}

\begin{equation}
\xi _{3}\equiv \frac{gE}{AD}\simeq \frac{gd_{eff}}{A^{\frac{2}{3}}DP^{^{%
\frac{1}{3}}}}  \label{VGL19}
\end{equation}%
The second equality in the expression for $\xi _{3}$ pertains to the limit $%
A\gg D$. It is again possible to introduce crossovers $\ $at $\xi _{3}\simeq
1$ and $L=L_{\ast }$ and so forth.

If one next increases the width $A$, the global persistence length given by
eqs (\ref{VGL24}) and (\ref{VGL27}) decreases rapidly to the value $P$ at $%
A=c_{3}P$ where $c_{3}$ is a numerical constant larger than unity. At the
same time the chain loses its local anisotropy: $\left\langle \left\vert
\sin \delta \right\vert \right\rangle =O\left( 1\right) $ (see eq (\ref{VGL31})). It is important
to realize that the chain remains confined to a thin slab $\left( D\lesssim
\pi P\right) $. Eq (\ref{VGL17}) with $g=P$ still holds though $\xi _{3}$ in
eq (\ref{VGL18}) is replaced by

\begin{equation}
\xi _{4}\equiv \frac{Pd_{eff}}{AD}  \label{VGL20}
\end{equation}

Ultimately, if we keep on increasing $A$, we attain the case where the chain
may be viewed as a two-dimensional pancake $\left( A> R_{e}\right) $.
Instead of eq (\ref{VGL17}), we have

\begin{equation}
\frac{F}{k_{B}T}\simeq \frac{R_{e}^{2}}{PL}+\frac{N_{\lambda }^{2}E}{%
R_{e}^{2}D}  \label{VGL21}
\end{equation}%
Minimization of $F$ with respect to $R_{e}$ yields

\begin{equation}
R_{e}\simeq \left( LP\right) ^{\frac{1}{2}}\left( \frac{Ld_{eff}}{PD}\right)
^{\frac{1}{4}}  \label{VGL22}
\end{equation}%
One recognizes the usual $3/4$ power law applicable to the excluded-volume
effect in two dimensions \cite{GEN2}. The excluded-volume parameter

\begin{equation}
Z\simeq \frac{Ld_{eff}}{PD}  \label{VGL23}
\end{equation}%
has to be greater than unity if eq (\ref{VGL23}) is to be valid otherwise $%
R\simeq \left( LP\right) ^{\frac{1}{2}}.$

The analysis given above has the drawback that the numerical coefficients
are unknown and may deviate substantially from unity. This is exemplified in
the application of eqs (\ref{VGL8}) and (\ref{VGL10}) to the recent
measurements on $\lambda$-phage DNA extended within nanochannels of
essentially square cross-sections \cite{REI1}. For instance, in the widest
channel of width $D=440$ nm, the dimensionless parameter $\xi _{1}$ is about
$0.11$ and $g=2.3$ $\mu$m which would lead to a crossover length $L_{\ast}$ of about $%
10$ $\mu$m (I have estimated $d_{eff}$ to be 4.6 nm on the basis of the
concentration of buffer used by Reisner et al). But we know that the
coefficients in eq (\ref{VGL2}) and (\ref{VGL3}) are somewhat smaller than
unity (see also ref \cite{ODI4}). In addition, there is a persistent finite
size effect for semiflexible chains which significantly suppresses the
excluded-volume interaction \cite{YAM2,MOO}. On the whole, $L_{\ast }$ could
be an order of magnitude larger so that the DNA of contour length $L=63$ $\mu$m
would only be slightly perturbed by excluded volume. This would explain why
the hairpin theory for the phantom worm \cite{ODI2} agrees well with the DNA extensions
\cite{REI1}. In a similar vein, Krishnan and Petrov \cite{KRI2} use the same
theory to explain the ionic-strength dependence of the DNA elongation
measured in some detail by Reisner et al \cite{REI2}. Theoretically, the strong dependence on salt then arises from the
exponential dependence of $g$ on the persistence length $P$ (see eq (\ref%
{VGL25})) \cite{KRI2}. The latter quantity was dealt with on an empirical
level by using the values from optical tweezer experiments \cite{BAU}. However, Reisner et al \cite%
{REI2} themselves argue that a blob picture could be valid on the basis of
assuming $g\equiv P$ (this would be regime 3 defined above). A reassessment of these experiments is warranted to reconcile the apparently opposing points of view.

As I argued above, the degree of orientational order depends on whether the
cross section of the channel in which the DNA is confined is square,
rectangular or slitlike of infinite extent. In the case of the $1000\times100$ nm$^{2}$ nanoslits used in ref. 12, the quantity $\xi _{3}=44$ from
eq (\ref{VGL19}) turns out to be very large because the ionic strength is
quite low ($d_{eff}=79$ nm). The DNA is predicted to be fully extended even
in 100 $\mu$m long slits as is indeed the case. Bonthuis at al \cite{BON}
have studied the radius of gyration of DNA in $2d$ slits as a function of
height $D$ right into the Odijk regime. There is only one abrupt transition
at $D\simeq 2P$ i.e. there are no intermediate regimes in agreement with the
analysis presented here.

In summary, a scaling analysis of nanoconfined DNA has been presented based
on orientational order imposed by the channel walls and a global persistence
length greatly enhanced by entropic depletion. Clearly, more experiments are
needed to delineate the regimes proposed here.

\section*{Acknowledgments}

I thank Peter Prinsen for logistic help and David C. Schwartz, Madhavi Krishnan, Walter Reisner, Douwe Jan Bonthuis and
Christine Meyer for discussions and correspondence.

\section*{Appendix I}

The global persistence length has been computed in the mechanical limit
for the hairpin configurations (see \cite{ODI2}; the bent DNA remains
double-stranded).

\begin{equation}
g=3.3082\hspace{2pt}\overline{r}\exp \left( F\left( \overline{r}\right) \right) /k_{B}T
\label{VGL24}
\end{equation}%
In the case of nanochannels of square cross section, the free energy of a
hairpin bend is given by

\begin{equation}
F_{s}\left( \overline{r}\right) /k_{B}T=\frac{E_{m}P}{\overline{r}}-3\ln \left( \frac{%
D-\overline{r}\sqrt{2}}{D}\right) -\ln \left( \frac{8}{3\pi }\right)
\label{VGL25}
\end{equation}%
with an optimum radius

\begin{equation}
\overline{r}=\frac{1}{6}\left[ \left( E_{m}^{2}P^{2}+6\sqrt{2}E_{m}DP\right) ^{%
\frac{1}{2}}-E_{m}P\right]  \label{VGL26}
\end{equation}%
($E_{m}=1.5071$). In the case of nanoslits, these
variables are

\begin{equation}
F_{\text{slit}}\left( \overline{r}\right) =\frac{E_{m}P}{\overline{r}}-\ln \left[ \left(
\frac{A-2\overline{r}}{A}\right) \frac{D}{\pi \overline{r}}\right] +1  \label{VGL27}
\end{equation}

\begin{equation}
\overline{r}=\frac{E_{m}PA}{A+2E_{m}P}  \label{VGL28}
\end{equation}%
As $A$ becomes very large, $g$ tends to $P$, at least to the leading order.
The limit is not precise because ultimately fluctuations cause the
mechanical approximation to break down \cite{ODI2}.

\section*{Appendix II}

To a good approximation, the orientation-translation distribution of the
long DNA is that of a chain confined in an appropriate harmonic well \cite%
{JO}. Integrating over the translational degrees of freedom, one is left
with a Gaussian distribution

\begin{equation}
f\sim \exp \left( -\frac{1}{2}G_{D}\theta _{x}^{2}\right) \exp \left( -\frac{%
1}{2}G_{A}\theta _{y}^{2}\right)  \label{VGL29}
\end{equation}%
for a fluctuating DNA segment (the nanoslit is $D$ wide in the $x$ direction
and $A$ wide in the $y$ direction; $G_{D}\simeq \left( P/D\right) ^{\frac{2}{%
3}}$ and $G_{A}\simeq \left( P/A\right) ^{\frac{2}{3}}$, see eq (\ref{VGL2}%
)). Hence, we have
\begin{widetext}
\begin{equation}
\left\langle \left\vert \sin \delta \right\vert \right\rangle
=\int\limits_{-\frac{1}{2}\pi }^{\frac{1}{2}\pi }d\theta _{x}\int\limits_{-%
\frac{1}{2}\pi }^{\frac{1}{2}\pi }d\theta _{y}\int\limits_{-\frac{1}{2}\pi
}^{\frac{1}{2}\pi }dw_{x}\int\limits_{-\frac{1}{2}\pi }^{\frac{1}{2}\pi
}dw_{y}f\left( \theta _{x},\theta _{y}\right) f\left( w_{x},w_{y}\right)
\left\vert \sin \delta\right\vert  \label{VGL30}
\end{equation}
\end{widetext}
(the appropriate normalization is included in eq (\ref{VGL30})). To the leading order
integrals like eq (\ref{VGL30}) are readily estimated \cite{ONS,ODI6}

\begin{equation}
\left\langle \left\vert \sin \delta \right\vert \right\rangle \simeq \left[
\frac{G_{D}+G_{A}+1}{\left( G_{D}+1\right) \left( G_{A}+1\right) }\right] ^{%
\frac{1}{2}}  \label{VGL31}
\end{equation}%
where the constant unity has been added to ensure that $\left\langle
\left\vert \sin \delta \right\vert \right\rangle =O\left( 1\right) $ in the
isotropic limit.

\end{document}